\documentclass[apj]{emulateapj}
\usepackage{graphicx}
\usepackage{amsmath}

\slugcomment{To appear in the {\it Astrophysical Journal}}
\shorttitle{Burn out or fade away?}
\shortauthors{J.J.~Drake et al.}
\begin{document}

\title{Burn Out or Fade Away? On the X-ray and Magnetic Death of Intermediate 
Mass Stars}

\author{Jeremy J.~Drake\altaffilmark{1},  Jonathan Braithwaite\altaffilmark{2},
Vinay Kashyap\altaffilmark{1}, H.~Moritz~G\"unther\altaffilmark{1}  and Nicholas J.~Wright\altaffilmark{1,3}}

\affil{$^1$Smithsonian Astrophysical Observatory,
MS-3, \\ 60 Garden Street, \\ Cambridge, MA 02138}
\affil{$^2$Argelander Institut f\"ur Astronomie, Auf dem H\"ugel 71, 53121 Bonn, Germany}
\affil{$^3$Centre for Astrophysics Research, STRI, University of Hertfordshire, College Lane Campus, Hatfield AL10 9AB, United Kingdom}
\email{jdrake@cfa.harvard.edu}

\begin{abstract}

The nature of the mechanisms apparently driving X-rays from intermediate mass stars lacking strong convection zones or massive winds remains poorly understood, and the possible role of hidden, lower mass close companions is still unclear.  A 20~ks {\it Chandra} HRC-I observation of HR~4796A, an 8 Myr old main sequence A0 star devoid of close stellar companions, has been used to search for a signature or remnant of magnetic activity from the Herbig Ae phase.  X-rays were not detected and the X-ray luminosity upper limit was $L_X\leq 1.3\times 10^{27}$~erg~s$^{-1}$.  The result is discussed in the context of various scenarios for generating magnetic activity, including rotational shear and subsurface convection.  A dynamo driven by natal differential rotation is unlikely to produce observable X rays, chiefly because of the difficulty in getting the dissipated energy up to the surface of the star.   A subsurface convection layer produced by the ionisation of helium could host a dynamo that should be effective throughout the main-sequence but can only produce X-ray luminosities of order $10^{25}$ erg s$^{-1}$.  This luminosity lies only moderately below the current detection limit for Vega.  Our study supports the idea that X-ray production in Herbig Ae/Be stars is linked largely to the accretion process rather than the properties of the underlying star, and that early A stars generally decline in X-ray luminosity at least 100,000 fold in only a few million years.  

\end{abstract}


\section{Introduction}
\label{s:intro}

The issue of whether late B and early A-type stars are capable of
generating and sustaining significant X-rays remains an 
outstanding problem in high energy stellar physics.  The problem is
relevant to both our understanding of the physics and structure of the
stars themselves and the evolution of their immediate environments.

There are two known ways non-accreting single stars are able to produce X-rays. In O and early B stars, shocks present in supersonic radiatively-driven winds produce heating. Stars of type G and later, on the other hand, have thick convective envelopes with magnetic dynamo activity, much of whose energy is dissipated in the corona \citep[e.g.][]{Vaiana.etal:81}. Thick convective envelopes are absent above about 1.5M$_\odot$, but thin surface convection layers present up to about type A5 appear to be enough to produce modest coronal heating and X ray emission \citep[e.g.][]{Robrade.Schmitt:10,Gunther.etal:12}.
Between these two regimes lie the early A- and late B-type stars, which are insufficiently luminous to produce massive radiatively-driven winds and also lack significant convection at the surface. Peculiar metal abundances are sometimes  observed in these stars, ``skin diseases'' produced by various chemical separation phenomena at the surfaces such as gravitational settling and radiative levitation, which in more active stars are smothered by mixing processes. This quietness appears also to manifest itself in X-ray quietness---indeed, various ROSAT searches for
X-ray emission from normal A-type stars have found most to be X-ray dark down to limits of
$L_X <  $~a few~$10^{27}$--$10^{28}$~erg~s$^{-1}$ \cite[e.g.][]{Simon.etal:95,Schroder.Schmitt:07},  with
significantly lower limits for nearby examples such as Vega. For this A0~V star, 
\cite{Pease.etal:06} obtained 
$L_X < 3\times 10^{25}$~erg~s$^{-1}$ from {\it Chandra} observations, corresponding to a bolometric fraction limit of $L_X/L_{bol}<9\times 10^{-11}$ \citep[see also][]{Ayres:08}.   
The spectral type limit earlier than which A-stars begin to be plausibly X-ray dark appears to be about A5 \citep[e.g.][]{Robrade.Schmitt:10,Gunther.etal:12}, corresponding to the spectral type of $\beta$~Pictoris from which \citet{Gunther.etal:12} recently confirmed the very weak X-ray emission tentatively identified by \citet{Hempel.etal:05}.  Evidence is also building that the magnetic A and late B stars might maintain significant X-ray output \citep[e.g.][]{Drake:98,Robrade.Schmitt:11,Stelzer.etal:11}.

While it is tempting to declare the late B and early A-type stars
essentially devoid of significant X-ray activity, a small fraction {\em are}
seen in positional coincidence with bright X-ray sources.  \citet{Schroder.Schmitt:07} found that 342 A-type stars in the Bright Star Catalogue \citep{Hoffleit.Jaschek:91} can potentially be associated with X-ray sources found in ROSAT surveys, corresponding to a detection rate of 10--15\%.   The question is what fraction of the X-ray detections are due to unseen late-type companions?   
In a sample of 11 late B-type main-sequence stars with resolved close companions at arcsecond separation---smaller than can be resolved by ROSAT---\citet{Stelzer.etal:06}  still found 7 of the B stars to be coincident with X-ray emission seen in {\it Chandra} observations capable of resolving the known components.  More recently, \citet{De_Rosa.etal:11} found from an adaptive optics survey that B6--A7 stars coincident with an X-ray source are three times more likely to have close companions than a sample with no corresponding X-ray detections.  This survey provides perhaps the strongest evidence to date that the source of the X-ray emission, at least in a sizable fraction of the sample, is the candidate companion.

One reason to suspect that at least some intermediate mass main-sequence stars are X-ray emitters is that pre-main
sequence Herbig Ae/Be (HAeBe) stars---the intermediate mass nearly fully radiative counterparts to
classical T~Tauri stars---are routinely found to be coincident with X-ray sources with 
luminosities of a few $10^{31}$~erg~s$^{-1}$ down to about $10^{29}$~erg~s$^{-1}$
\citep{Damiani.etal:94,Zinnecker.Preibisch:94,Hamaguchi.etal:05,
Stelzer.etal:06,Stelzer.etal:09,Hamidouche.etal:08}.   \citet{Stelzer.etal:06} found an
overall detection fraction of 76\%\ for a sample of 17 HAeBes, and 
only half of these have known unresolved
companions.  They found that the observed X-ray emission cannot be explained by known companion stars in 35\%\ of the sample.   If the Herbig~Ae/Be stars are indeed responsible for some of these detections,  they are more vigorous sources than any true X-ray emitters among their more evolved main-sequence siblings.  
\citet{Stelzer.etal:09} noted that if HAeBe X-rays are simply indicative of coronally active low-mass companions, their detection statistics imply a high fraction of higher order multiple systems among Herbig stars.

Since the limits to the X-ray luminosity of stars like Vega (A0~V),
with an age likely in the range of  about $100$--500~Myr \citep{Barrado_y_Navascues:98,Hill.etal:10,Yoon.etal:10},
are currently quite stringent at $L_X\la 10^{25}$~erg~s$^{-1}$, 
any X-ray emission it produced in a pre-main sequence star phase must have declined rapidly as it 
evolved to the main-sequence---by perhaps  6 orders of
magnitude or more.   Any X-ray activity of intermediate mass stars then occurs either particularly in the HAeBe phase and is related to the presence of disks, accretion or associated jet-type activity, as appears to be the case for HD163296 \citep[e.g.][]{Swartz.etal:05,Gunther.Schmitt:09}, or else represents magnetic dissipation and decays due to a limited non-thermal energy reservoir as originally proposed in the primordial rotational sheer dynamo model of \citet[][see also \citealt{Spruit:02,Braithwaite:06}]{Tout.Pringle:95}.   Spectropolarimetric time series of Herbig Ae stars provide some support for the latter, in revealing evidence for dipolar surface fields with polar magnetic field strengths of up to several hundred Gauss \citep[e.g.][]{Hubrig.etal:14}.

Arguably the most important aspect of any observational campaign hoping to resolve the question of X-ray activity in intermediate mass stars is to disentangle the role of companions.  Studies of early A-type stars in whose single nature we have a high degree of confidence are therefore of special value.  HR~4976A is one of these.  It is a rare example of a nearby ($72.8 \pm 1.7$~pc \citealt{van_Leeuwen:07}) very young main-sequence A0 star, with an age estimated to be about 8~Myr.  Here we report on {\it Chandra} observations of HR~4976A obtained to search for evidence of any remaining X-ray activity.

\section{HR 4796A}

HR~4976A rose to prominence following the discovery of its remnant dusty disk based on {\it Infrared Astronomical Satellite} data \citep{Jura:91}, and through subsequent ground-based near-IR imaging \citep{Koerner.etal:98,Jayawardhana.etal:98}.  Its disk was later  imaged at high spatial resolution by the {\it Hubble Space Telescope} \citep{Schneider.etal:99,Schneider.etal:09} and by ground-base adaptive optics \citep{Thalmann.etal:11}.  Imaging reveals signs of asymmetry and clearing that could be signatures of planetary mass perturbers \citep{Wyatt.etal:99,Schneider.etal:09,Thalmann.etal:11} but rules out close stellar mass companions.

Of key importance to the questions of its possible X-ray activity is the age of HR~4796A.  A nearby M dwarf with the same proper motion is very likely a binary companion and 
enabled \citet{Stauffer.etal:95} to assess the system age as $8\pm 2$~Myr based on isochrone fitting and Li abundance.  This estimate was revised to 10~Myr by \citet{Jura.etal:98} based on the {\it Hipparcos} distance, and by 
\citet{Jayawardhana.etal:98}, who also incorporated more recent photometry and bolometric corrections and assessed $8\pm 3$~Myr.  \citet{Jayawardhana.etal:98} noted that the isolated location of HR~4796, free from molecular clouds or substantial dust extinction, provides a lower age limit of a few Myr.  \citet{Stauffer.etal:95} based an upper limit to the age of 9-11~Myr on the strong Li absorption of HR~4796B.   

Perhaps some caution should be applied to these age interpretations though.  \citet{Tout.etal:99} have shown that the accretion history of  pre-main-sequence stars can affect their positions in the HR diagram, inducing potential age errors from isochrone fitting.  The Li abundance can also be affected by accretion.  \citet{Soderblom.etal:13} also point out that published stellar model isochrones for young stars can be significantly different from one another.  While HR~4976 has been identified as a member of the TW~Hya association (TWA 11; \citealt{Webb.etal:99}), whose age is often thought well-established at 8-10~Myr, \citet{Weinberger.etal:13} have pointed out an apparent age spread of several Myr among association members.  
In the light of such potential uncertainties, we allow conservatively the age of HR~4976 to be up to a factor of two higher than its nominal value, and adopt an age range of 5--16~Myr.

\section{Observations and Analysis}
\label{s:anal}

{\it Chandra} observed HR~4796 using the HRC-I detector 
on 2006 December 2 with a net exposure  of 20,800~ks.  The HRC-I was preferred over ACIS because of its greater effective area at energies $E\la 0.5$~keV where a low-activity corona of cooler, more solar-like temperatures of $T\approx 2\times 10^{6}$~K might show up.

We reduced the HRC-I data with CIAO 4.2 and applied detector pulse height-based background filtering\footnote{http://cxc.harvard.edu/ciao/threads/hrci\_bg\_spectra/ } .
Conservatively, this removes at most $5\%$ of the source
counts
and we have included this correction in all the count rate and flux 
estimates below.  The observation does not exhibit significant 
background flaring, and thus we did not apply additional
time filtering.  

\begin{figure}
\begin{center}
\includegraphics[width=0.45\textwidth]{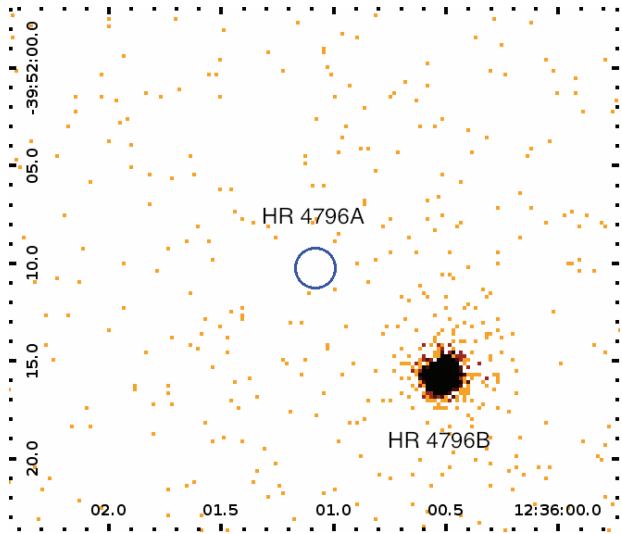}
\end{center}
\caption{
The HRC-I image in the vicinity of HR~4796A from the 20.8~ks HRC-I  observation (ObsID 7414) described here.  The main target was not detected, with no events lying in a region of radius $1\arcsec$ centered on the position of the star.  A large region was used to get an accurate assessment of the background with which to determine a count rate upper limit of $2.8\times 10^{-4}$ 
count~s$^{-1}$.  The bright source to the south-west is HR~4796B, an M2.5V star used by \citet{Stauffer.etal:95} to estimate an age of $8\pm2$~Myr for the system.
}
\label{f:sky}
\end{figure}

The resulting X-ray image of the sky in the region around HR~4796 is illustrated in Figure~\ref{f:sky}.
Close visual inspection of the data in the vicinity of HR~4796A revealed no tangible signs of an X-ray source, or of any events within $1\arcsec$ of the stellar position.  We therefore proceeded to obtain an estimate of the upper limit to the X-ray flux.

We defined a large rectangular region offset from the
putative location of HR\,4796A to measure the background,
and found that the expected background under the source in a circular region of
$1\arcsec$~radius that encloses $97\%$ of the point spread function (PSF), 
is $1.86\pm0.07$.  
Adopting the formulism of 
\citet{van_Dyk.etal:01}, the nominal
$68\%$ credible range on the source intensity is
$[0,7.4\times10^{-5}]$~ct~s$^{-1}$. 
However, because the source is undetected,
we can also estimate the intensity that would be required
for an unambiguous detection \citep[see][]{Kashyap.etal:10}: for a source 
to be detected at a significance of
at least $99.7\%$ (corresponding to a Gaussian-equivalent
``$3\sigma$'' detection with 7 counts) with a probability of
$0.5$, its intensity must be at least $2.8\times10^{-4}$~ct~s$^{-1}$,
and the lack of such a detection places this upper limit
on the intensity of HR\,4796A.

\begin{figure}
\begin{center}
\includegraphics[width=0.49\textwidth]{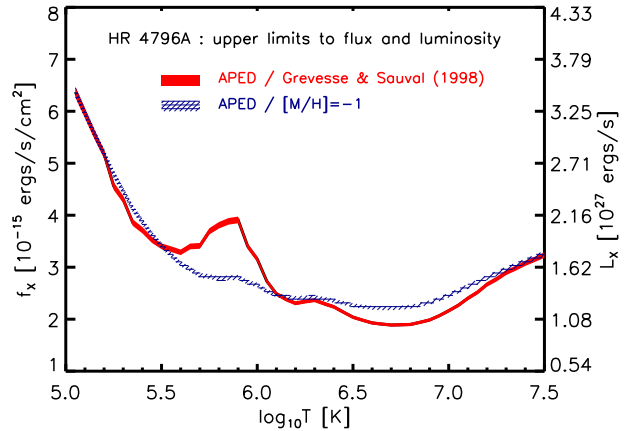}
\end{center}
\caption{
The relationship between source count rate and X-ray flux and luminosity as a function of temperature for HR~4796A for its distance of 73~pc.  Curves were computed using the APED optically-thin plasma radiative loss model.  Here we adopt the luminosity limit at $\log T=6.3$, $L_X\leq 1.3\times 10^{27}$~erg~s$^{-1}$, which is also independent of the assumed metallicity.
}
\label{f:lxvst}
\end{figure}

The count rate upper limit was converted to flux and luminosity upper limits by folding the response of the instrument with the radiative loss function for an isothermal plasma at temperatures in the range $10^{5}$Ð-$10^{7}$~K.  For the latter, we adopted the APED radiative loss model \citep{Smith.etal:01} as implemented in the PINTofALE\footnote{The Package for Interactive Analysis of Line Emission, freely available from http://hea-www.harvard.edu/PINTofALE/ } IDL\footnote{Interactive Data Language, Research Systems Inc.}  software  \citep{Kashyap.Drake:00}
and assumed the solar abundance mixture of \citet{Grevesse.Sauval:98}.   This X-ray flux and luminosity  upper limit locus (assuming a distance of 73~pc) is illustrated, together with one computed with metal abundances reduced by a factor of ten, in Figure~\ref{f:lxvst}.   It is difficult to know what the characteristic temperature of X-ray emission from HR~4796A might be, but since coronal temperatures in main-sequence stars with weak magnetic activity, such as the Sun, are generally about 1--$2\times 10^6$~K, 
for the ensuing discussion here we adopt the luminosity limit at $\log T=6.3$, which is also independent of the assumed metallicity.  The resulting X-ray luminosity for HR~4796A is $L_X\leq 1.3\times 10^{27}$~erg~s$^{-1}$, corresponding to an X-ray to bolometric luminosity ratio limit of  $L_X/L_{bol}\leq 1.9\times 10^{-8}$ based on the bolometric luminosity of HR\,4796A of $L_{bol}=18.1 L_\odot$ \citep{Koerner.etal:98}.

\section{Discussion}
\label{s:discuss}

While our {\it Chandra} observations have resulted only in an upper limit to the X-ray luminosity of HR~4796A, it is of interest to place this limit in the context of possible mechanisms for generating X-rays in radiative stars.

There have been two different propositions 
explaining X-ray activity of HAeBe
stars and its evident precipitate decline.  The first is that magnetic
activity depends on star-disk interactions and, perhaps, associated jet activity \citep[see, e.g.,][]{Gunther.Schmitt:09}; once the accreting gas disk is
sufficiently dissipated the magnetic activity is curtailed
\citep{Hamaguchi.etal:05}.  The second is the gradual dissipation by means of some magnetic dynamo of
primordial rotational  shear that remains between surface and deeper layers of the star from its initial collapse 
(\citealt{Vigneron.etal:90,Tout.Pringle:95,Spruit:02}).   We consider the latter below, together with two other potential mechanisms of X-ray generation: decay of a magnetic field left over from formation, and a subsurface convection zone thought to exist in late B-type and early A-type stars.

\subsection{X-rays from a shear dynamo}
\label{s:shear}

In this section we describe two mechanisms which tap the energy in the primordial differential rotation and potentially result in X-ray emission.  Both are dynamos consisting of two steps: first, any seed magnetic field present in the star will be wound up by the differential rotation, to become a predominantly toroidal field. Then, to close the dynamo loop, it is necessary to convert some of this toroidal component back into poloidal field which can be wound up again.  In convective dynamos, this conversion from toroidal into poloidal field is achieved by convective motion; in a radiative zone some magnetic instability can be used.  
\citet{Tout.Pringle:95} developed a dynamo model in which the toroidal-into-poloidal conversion works via the Parker \citep[magnetic buoyancy;][]{Parker:66} instability. \citet{Spruit:02} and \citet{Braithwaite:06} developed a similar idea, though in more physical detail and based on a different instability, the so-called Tayler instability \citep{Tayler:73}.  As we shall see, the two different treatments result in what are at face value very different timescales for dissipation of shear energy, though there are large uncertainties in both.

The relations of \citet{Malkov:07} indicate a mass of $2.6\,M_\odot$ for spectral type A0.  
Following Eqn.~2.6 of \citet{Tout.Pringle:95}, the available shear energy for a star of this mass could be as high as $7\times 10^{47}$~erg, and, as those authors pointed out, sufficient to power a typical X-ray luminosity of HAeBe stars for the stellar lifetime even with a conversion efficiency of a percent or less.
While the
parameters are somewhat uncertain, \citet{Tout.Pringle:95} found the initial rotational shear, $\Delta\Omega_0$, to decline with time, $t$, according to 
\begin{equation}\label{tp_decay}
\Delta\Omega=\frac{\Delta\Omega_0}{\left(1+t/\tau\right)},
\end{equation}
where $\tau$ is the decay timescale ($t_0$ in the \citealt{Tout.Pringle:95} nomenclature), and the 
initial X-ray luminosity, $L_{X_0}$,  to decay in time as
\begin{equation}
L_X=\frac{L_{X_0}}{(1+t/\tau)^{3}}. 
\label{e:lxtp}
\end{equation}
\cite{Tout.Pringle:95} found $\tau\sim 10^6$~yr, based on reasonable guesses of relevant parameters.  This also corresponds to the timescale of evolution of HAeBe stars and the typical time over which they appear X-ray bright \citep[e.g.][]{Hamaguchi.etal:05}. 
The parameter describing the efficiency of field generation on which the timescale $\tau$ depends to the inverse third power lacks empirical constraints, however. Since its value can only be considered an order of magnitude estimate, the decay timescale should also be considered very uncertain, by perhaps up to a factor of a thousand or so.  

\citet[][see also \citealt{Spruit:99,Braithwaite:06}]{Spruit:02}  noted that the Tayler instability \citep{Tayler:73} is likely to be more relevant for a stably-stratified star than than the Parker instability since it sets in at much lower magnetic field strengths. This is an interchange instability of toroidal fields, similar to the pinch instability in plasma physics. 
The instability necessarily involves some movement in the radial direction, and so has to work against the stable stratification. In the case where the composition gradients are insignificant (as in young stars), thermal diffusion can facilitate radial motions by reducing the effect of the stratification. In this regime, Spruit obtained an expression (Eqn.~29 of \citealt{Spruit:02}) of the following type for the 
azimuthal stress due to the field generated by the dynamo for the case of negligible compositional gradient,
\begin{equation}
S\approx \frac{B_r B_\phi}{4\pi} = \Lambda \Omega^{3/2}\Delta \Omega,
\end{equation}
where $\Omega$ is the rotation rate of the layers under consideration.  Here we have included general terms related to the structure of the star in question into a parameter $\Lambda$, and for simplicity of comparison have adopted the nomenclature of \citet{Tout.Pringle:95} in which $\Delta\Omega$ is the change in angular velocity over the region of the star in which the dynamo is expected to operate. 
This shear is continually eroded by the azimuthal stress and we can write 
\begin{equation}
\frac{d(\Delta\Omega)}{dt} \propto \Omega^{3/2}\Delta \Omega.
\label{e:dodt}
\end{equation}
The decay of shear is then likely to depend on whether significant angular momentum is lost and the average rotation rate in the dynamo region declines with time, or angular momentum is conserved and the average rotation rate can be approximated as being constant.
During the HAeBe phase in which angular momentum can be dissipated by a wind, and assuming $\Delta\Omega \propto \Omega$, Eqn.~\ref{e:dodt} leads to the solution 
\begin{equation}
\Delta\Omega=\frac{\Delta\Omega_0}{\left(1+t/\tau\right)^{2/3}},
\end{equation}
where the timescale $\tau$ is the decay timescale, as before; the time taken for the shear to drop to half its original value is $1.8\tau$. 
 
For an early A-type star lacking a strong wind and significant magnetic braking, the total angular momentum should remain approximately constant.  If we assume in this case that the mean rotation velocity of the dynamo layers is relatively unchanged by the dissipation of shear and $\Omega$ is approximately constant, the solution to Eqn.~\ref{e:dodt} is simply $\Delta \Omega=\Delta\Omega_0 \exp (-t/\tau)$,where $\Delta\Omega_0$ is the initial rotational shear and $\tau$ is the timescale for its decay. 
This time dependence differs from that of \citet[][Eqn.~3.13]{Tout.Pringle:95} and Eqn (\ref{tp_decay}) here, who obtained a $1/t$ rather than exponential decay.  

In order to understand the time dependence of surface X-ray emission for a Tayler-Spruit dynamo, we would need to know the rate at which magnetic field is brought to the surface.  
\citet{Mullan.MacDonald:05} argued that the Tayler and buoyancy instabilities are not mutually exclusive and that fields generated on the basis of the Tayler instability can be brought to the surface by magnetic buoyancy; we return to this further below.  Here, while the following approach is somewhat arbitrary, for the sake of comparison with the \citet{Tout.Pringle:95} result we apply the same arguments used by them to estimate surface X-ray activity.

We assume some fraction, $\varepsilon$, of the magnetic field arriving at the surface by upward buoyant drift  is scavenged off eventually to be dissipated in the form of X-rays.  In analogy with  \citet{Tout.Pringle:92}, the X-ray luminosity is then
\begin{equation}
L_X\approx \varepsilon\frac{4}{3}\pi R^3 \frac{d}{dt}\frac{B^2}{8\pi},
\end{equation}
where $R$  is the stellar radius.
\citet{Tout.Pringle:92} assumed that magnetic flux escapes at a speed of about $0.1v_A$, where $v_A$ is the average Alfv\'en speed.  For $v_A=B/\sqrt{4\pi\rho}$ and a timescale for the emergence of field of $R/0.1v_A$, we can write
\begin{equation}
L_X \approx  \varepsilon\frac{4}{3}\pi R^3 \frac{B^2}{8\pi} \frac{0.1v_A}{R} \approx   \frac{0.1 \varepsilon R^2}{12(\pi \rho)^{1/2}}B^3.
\label{e:lx_b}
\end{equation}
Since all the terms except for $B$ in Eqn.~\ref{e:lx_b} are constants for a given star, the X-ray luminosity then scales as the cube of the mean magnetic field. \citet{Spruit:02} finds that the azimuthal field should be much larger than the radial field, and we can ignore the latter in considering both the magnetic energy and Alfv\'en speed.  From Eqn.~22 of \citet{Spruit:02}, the azimuthal field depends only on the rotation rate, the differential rotation and quantities that are constants for a given star, $B_\phi \propto \Delta\Omega^{1/2}\Omega^{5/8}$.  If the average rotation velocity remains approximately constant for the main-sequence phase,  the X-ray luminosity is expected to scale as 
\begin{equation}
L_X \propto \Delta\Omega^{3/2}\Omega^{15/8} \propto \Delta\Omega^{3/2}. 
\end{equation}
Combining this with the solution to Eqn.~\ref{e:dodt}, the time dependence of the X-ray luminosity is then
\begin{equation}
L_X(t)=L_{X_0}\exp \left( - \frac{3t}{2 \tau}\right).
\label{e:lxsp}
\end{equation}

It would now be useful to estimate the timescale of the decay. 
From \citet[][Eqn.~32]{Spruit:02}, one can express the shear stress in terms of an effective viscosity, which is given by $\nu \sim r^2 \Omega \left(\frac{\Omega}{N}\right)^{1/2} \left(\frac{\kappa}{r^2N}\right)^{1/2}$, where $\kappa$ is the thermal diffusivity, $N$ is the buoyancy frequency and $r$ the radial coordinate. The timescale $\tau$ for the dissipation of the energy decaying as $E_{dr}=E_{dr}(t=0) \exp(-t/\tau)$ is then given by
\begin{eqnarray}
\tau&\sim & \frac{r^2}{\nu} \sim \Omega^{-1} \left(\frac{N}{\Omega}\right)^{1/2} \left(\frac{R^2N}{\kappa}\right)^{1/2},\\
& \sim & \frac{P_{\rm rot}^{3/2}\tau_{\rm KH}^{1/2}}{P_{\rm bu}},
\end{eqnarray}
where $N \sim \Omega$ close to the break-up velocity, 
$P_{rot}$ and $P_{\rm bu}$ are the spin period and break-up spin period, $\tau_{\rm KH}$ is the Kelvin-Helmholtz timescale.
 A $2.6 M_\odot$ star has $\tau_{\rm KH}\sim2\times10^6$ years and $P_{\rm bu}\sim 5$ hours, so with a rotation period of $1$ day we would have a decay timescale of $300$ years.  Thus, for this formalism, the differential rotation is probably damped too quickly to see X-rays by the time the star becomes observable. However, this timescale assumes that the dynamo is at saturation, whereas in reality it may take rather longer than this timescale just to reach saturation. This though should in any case happen faster than $\tau_{\rm KH}$, so this mechanism should dissipate shear energy on a timescale shorter than the thermal timescale on which pre-MS stars evolve (see Braithwaite \& Spruit 2014). Consequently, we can expect it to work continuously and efficiently as the convective envelope retreats outwards and to keep the radiative zones in approximately solid-body rotation at all times. In this picture, it dissipates the shear energy present in the convective zones at a rate at least a couple of orders of magnitude smaller than the luminosity of the pre-MS star.  

Since the dynamo field strength tends to zero towards the surface, in order to dissipate some of the magnetic energy in X-ray emission it is necessary to bring stronger fields from further down upwards to the surface.  This ultimately determines the efficiency with which the differential-rotation energy is converted to X-rays.   Bringing field to the surface is difficult because it has to be brought from deep in the envelope, where diffusive buoyancy acts slowly  \cite[see, e.g.,][]{MacGregor.Cassinelli:03}.  As we noted and employed above,  \citet{Tout.Pringle:95} assumed a buoyancy turnover velocity of $0.1v_A$.  \citet{Mullan.MacDonald:05} looked at differentially-rotating radiative stars with the Tayler-Spruit mechanism and found that the fields produced can be brought to the surface by buoyancy instability, which operates on a dynamic timescale. They find that field strengths of order $100$~G are possible at the surface.  Such a field strength is compatible with spectropolarimetric observations of Herbig~Ae stars that find dipole-like fields with polar field strengths of up to a few hundred G \citep[e.g.][]{Hubrig.etal:14}.
The energy dissipation rate is more difficult to calculate, since it depends on the timescale and frequency with which magnetic features are brought upwards, but we can estimate this timescale as simply the Alfv\'en timescale for fields of this strength, which is about 1~yr. This gives a rate of energy deposition through the photosphere of around $10^{28}$ erg s$^{-1}$, but, as noted above, this would be expected to die away on a rather short timescale.

This scenario is, then, much more pessimistic for the detection of X-rays on the early main sequence than the \citet{Tout.Pringle:95} picture, although the large uncertainties in both methods for estimating the decay timescale cannot be overemphasized.

\subsection{Decay of primordial magnetic fields}
\label{s:fossil}

According to the failed fossil theory \citep{Braithwaite.Cantiello:13}, radiative stars can host weak magnetic fields in continuous dynamic evolution. The star could inherit a field from the parent cloud from which it formed or a field could be left behind by a pre-main-sequence convective dynamo. Either way, when the star is formed, its magnetic field is not in MHD equilibrium, and evolves on its own dynamic timescale. As it does so, magnetic energy is lost and the field strength drops. Whilst in the strongly-magnetic Ap stars an equilibrium is quickly reached and the field essentially stops evolving (a fossil field), a so-called {\it failed fossil} field is still evolving on a timescale given in terms of the Alfv\'en timescale and rotation rate by $\tau_{\rm A}^2 \Omega$, which is then equal to the age of the star $t$. 
 $E$ is the magnetic energy in the star and can be expressed in terms of an average field strength as $E\sim (4\pi/3)R^3 B^2/8\pi$. If $M$ is the mass of the star and $L$ is the characteristic length scale of the magnetic field, the Alfv\'en timescale is $\tau_{\rm A}=L/v_{\rm A}=L\sqrt{4\pi\rho}/B$ where the mean density $\rho=M/(4\pi/3)R^3$. Putting these together we find that the magnetic energy falls at a rate 
\begin{equation}
\dot{E} \sim -\frac{ML^2\Omega}{2t^2}.
\end{equation}
  Some fraction $\zeta$ of this energy will be dissipated at and above the stellar surface and give rise to X-ray emission. However, putting in the numbers we have:
\begin{align}
L_{\rm X} & =  -\zeta \frac{{\rm d}E}{{\rm d}t} \nonumber \\
  & \sim  4\cdot 10^{23}\, \zeta \left(\frac{M}{M_\odot}\right) 
\left(\frac{L}{R}\right)^2 \left(\frac{P}{\rm day}\right)^{-1} \left(\frac{t}{\rm Myr}\right)^{-2} {\rm erg\, s}^{-1}.
\end{align}
Clearly, magnetic fields of this kind are unlikely to produce observable X-ray emission, even in very young main-sequence stars.

In pre-main-sequence stars, the situation is rather different. In general, any strong magnetic field present in a convective star is quickly brought to the surface by its own buoyancy \citep{Braithwaite:12}, where presumably a significant fraction of its energy is dissipated in a hot corona. However, this buoyant expulsion happens on a short timescale, so that in accreting protostars, the result of the buoyancy is more likely to be the prevention of flux accretion. The X-ray flux from the dissipation of magnetic energy would then be indistinguishable from that from the accretion itself, except in certain cirumstances when spectroscopic signatures of accretion shocks might be observed \citep[e.g.][]{Kastner.etal:02,Stelzer.Schmitt:04,Drake:05,Gunther.etal:06,Brickhouse.etal:10}. A convective dynamo would then be responsible for X rays seen after accretion has ceased.

\subsection{Subsurface convection}
\label{s:conv}

More promising for the production of X-rays are subsurface convective layers (\citealt{Cantiello.Braithwaite:11}; Cantiello et al.\ in prep.). These layers, which owe their existence to opacity bumps from ionization of iron (in O and late B stars) and helium (in late B and early A stars), lie just below the stellar surface, in contrast to the photospheric convection in stars later than about A5.

Hosting a dynamo, the subsurface convection in early A and late B stars produces magnetic field. It is worth noting though that the convection in these layers transports a fraction of only about $10^{-3}$ of the stellar luminosity, so that the magnetic energy production will be correspondingly lower than that produced by the strong convection in a solar-type star. In any case, the magnetic field produced easily floats upwards through the overlying radiative layer. This works essentially because the magnetic field produces pressure without mass; magnetic features are in total pressure equilibrium with their surroundings and therefore have a lower gas pressure. Heat diffuses efficiently into them and so they maintain the same temperature as their surrounding, have a lower density and rise. This buoyant rise is limited by aerodynamic drag and takes place at the Alfv\'en speed.  Once the magnetic field crosses into the low-density environment above the photosphere, its energy is dissipated.  In solar-type and low-mass stars we see that X-ray emission is correlated with rotation speed up to a certain saturation level \citep[e.g.][]{Pizzolato.etal:03,Wright.etal:11}, when the Rossby number (defined as $P_{\rm rot}/\tau_{\rm conv}$) is around 0.1; at Rossby numbers below this the X-ray emission $L_{\rm X}$ accounts for a fraction of around $10^{-3}$ of the total luminosity $L_{\rm bol}$. The convective turnover time in an A0 star should be a few hours, so the Rossby number in a fast rotator will be $\ge 3$, which in solar-type stars would give $L_{\rm X}/L_{\rm bol}\approx 10^{-6}$. Obviously it is little better than speculation that a thin-layer dynamo would work in the same way in this respect as a dynamo operating throughout a convective envelope, but assuming it does, we have
\begin{equation}
\label{e:lx_conv}
L_{\rm X} \sim 10^{25} \left(\frac{P}{\rm 12 hr}\right)^{-2} {\rm erg\, s}^{-1}
\end{equation}
for a $2M_\odot$ star with $L=25L_\odot$, taking the approximate power of $-2$ on the period from \citet{Pizzolato.etal:03}, taking into account the aforementioned convective transport fraction of $10^{-3}$ and assuming an efficiency of $0.1$ in getting the magnetic energy through the overlying radiative layer, a rough estimate which comes from the density ratio (and consequent magnetic field strength ratio) across the radiative layer.

\subsection{Confronting theory and observations}

\begin{figure*}
\begin{center}
\includegraphics{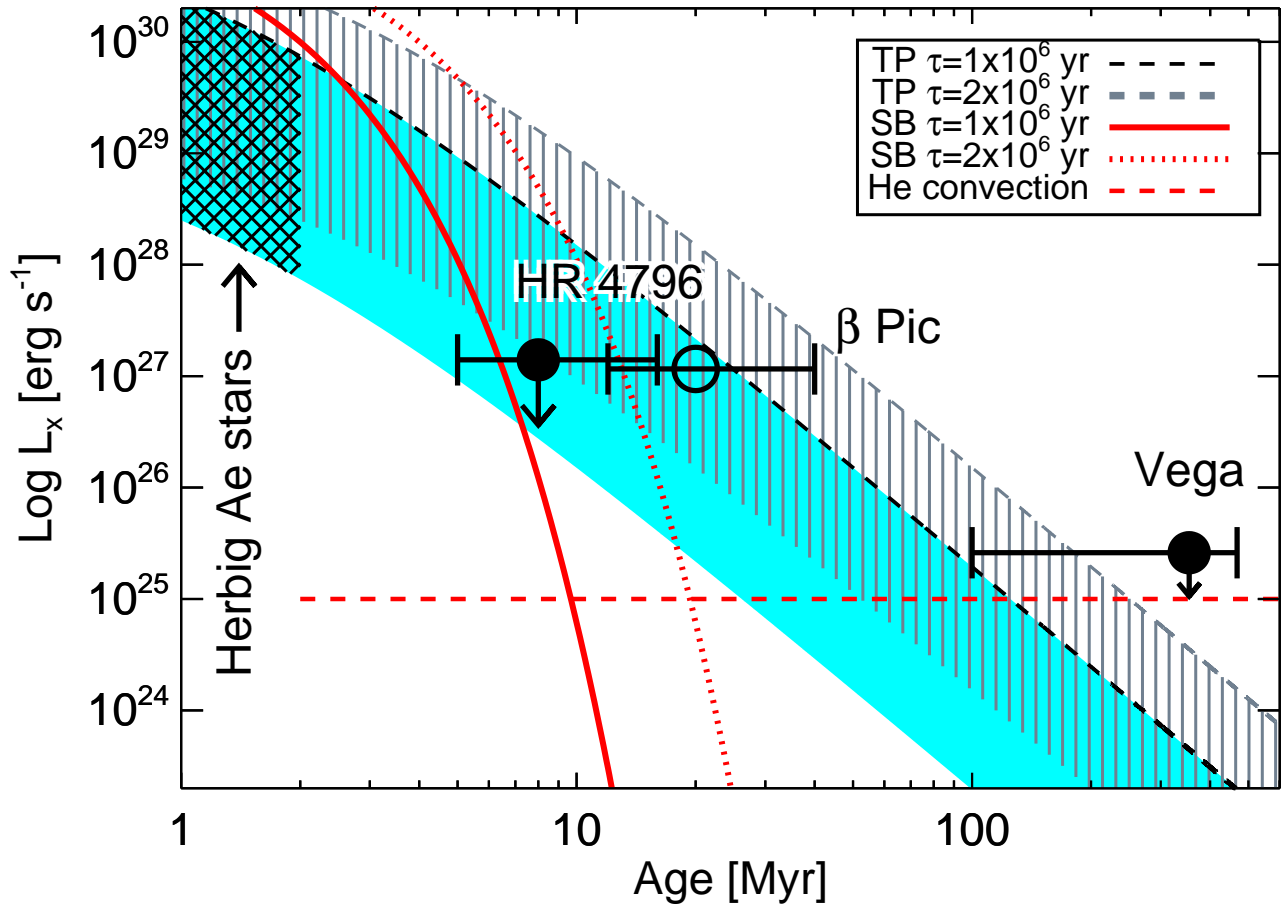}
\end{center}
\caption{X-ray luminosity vs.\ time for shear dynamo models compared with observations of single A-type stars.   The hashed region to the left represents the range of
 X-ray luminosities of Herbig~Ae stars.  The dashed curves correspond to a natal X-ray luminosity of $L_{X_0}=2\times 10^{31}$~erg~s$^{-1}$ and two different values for the rotation shear decay timescale for the  \citet[][TP]{Tout.Pringle:95} model.  The solid and dotted curves corresponds to an exponential X-ray luminosity decay that we infer from the rotational shear dissipation found by \citet[][SB; see also \citealt{Braithwaite:06}]{Spruit:02} assuming the \citet{Tout.Pringle:95} prescription for magnetic flux rising due to buoyancy.  
 The  horizontal dashed curve corresponds to our estimate of a base level X-ray luminosity due to a weak sub-surface convection zone.  The $\beta$~Pic detection of \citet{Gunther.etal:12} is shown with a hollow symbol because its X-ray emission likely originates from a thin surface convection zone.
\label{f:toutpringle}
}
\end{figure*}

The X-ray luminosity upper limit we obtained for HR~4796A is illustrated, together with that of Vega,  in the context of the 
X-ray production mechanisms discussed in \S\ref{s:shear}--\ref{s:conv}
in Figure~\ref{f:toutpringle}.  A similar figure was presented by \citet{Pease.etal:06} based on the stringent upper limit obtained for the X-ray luminosity of Vega.  Since that time, an analysis by \citet{Yoon.etal:10} of high-resolution spectra, visible/near-IR fluxes and optical interferometry suggest Vega is significantly less massive and older than had been thought, with an age of 
$454\pm13$~Myr, instead of the $200\pm 100$~Myr estimate based on kinematic similarity with, and assumed membership of, the Castor moving group \citep{Barrado_y_Navascues:98}.
A greater age for Vega would imply a less stringent constraint on shear and failed fossil dynamo mechanisms, but at the same time emphasises the value of HR~4796A as a constraint at young ages.

The hashed region to the left in Figure~\ref{f:toutpringle} represents the observed range of X-ray luminosity of Herbig Ae stars, and the shaded continuation to the right that which might be expected of main-sequence early A-type stars for rotational shear decay timescales of $10^6$ and $2\times 10^6$~yr .  They are bounded by initial X-ray luminosities $L_{X_0}=2\times 10^{29}$--$2\times 10^{31}$~erg~s$^{-1}$, which match the approximate luminosity range of Herbig AeBe stars \citep[e.g.][]{Damiani.etal:94,Zinnecker.Preibisch:94,Hamaguchi.etal:05,Stelzer.etal:06}.   

\citet{Skinner.etal:04} used the \citet{Tout.Pringle:95} formula to predict 
the present-day $L_X$ of the closest known HAeBe star, HD~104237:  while the estimate 
appears too high by a factor of about 4--20, within the
uncertainties in the age and spectral type of HD~104237 the disagreement is perhaps not unreasonable as  \citet{Skinner.etal:04} note.  
Observations of typical HAeBe stars
provide some additional calibration: 
\citep{Hamaguchi.etal:05} found evidence that the age of X-ray activity decay is $10^6$~yr.  
At an age of $>100$~Myr and assuming the decay law remains applicable to greater ages, the X-ray luminosity of Vega is then
expected to have decayed by 7 orders of magnitude to an unobservable
$10^{24}$~erg~cm$^{-2}$~s$^{-1}$ or so, in agreement with current limits
\citep{Pease.etal:06}.  In contrast, very young early-type A stars at ages of
about 10~Myr should have declined in $L_X$ by factors of 
$10^3$-$10^4$.  

For an A0 star of mass $2.6\,M_\odot$, the Tout-Pringle initial X-ray luminosity is  $L_{X_0}\approx 10^{31}$~erg~s$^{-1}$, or close to the case of the loci in Figure~\ref{f:toutpringle}.   At face value, our observed limit for HR~4976A weakly excludes the \citet{Tout.Pringle:95} model but is more consistent with our \citet{Spruit:02}-based formula assuming, probably optimistically, that magnetic field is brought efficiently to the surface by magnetic buoyancy.
However, as we have noted, both theoretical approaches involve significant uncertainty.  As \citet{Tout.Pringle:95} admit, their formulism includes some parameters whose values cannot be precisely defined or determined.
 These include physical aspects of the model that might be expected to be the same for all stars, such as the efficiency of conversion of magnetic flux into observable X-rays, and a star-specific parameter describing the natal rotational shear.   Within the Tout-Pringle framework, the observed spread in X-ray emission for stars at a given age and similar mass must in fact derive from the natal differential rotation.  

The observed rotation speed of HR~4796A is $v\sin i=152$~km~s$^{-1}$ \citep{Royer.etal:02}, the radius is $1.6 R_\odot$ \citep{Rhee.etal:07}, and the disk inclination is $76\deg$ \citep{Schneider.etal:09,Thalmann.etal:11}.  If the stellar and disk inclinations are the same, the stellar equatorial velocity is then 157~km~s$^{-1}$ and the rotation period is about 0.5 days.  In relation to the extensive sample of rotation velocities of \citet{Zorec.Royer:12}, HR~4796A lies near the middle of the distribution for stars with masses 2.4--$3.85 M_\odot$ and is rotationally typical.
In the presence of a shear-related spread in a Tout-Pringle type model, and not knowing what rotational shear, if any, remains in HR~4796A, our X-ray luminosity limit is only able to rule out shear dynamo decay timescales $\tau\ga 2\times 10^6$~yr.  

A fairly short timescale for dissipation of natal rotational shear is consistent with stellar rotation observations.  
From an extensive survey of stellar rotation rates, 
\citet{Zorec.Royer:12} have shown that the distribution of surface rotation periods for early A stars evolves only very slowly, over a large fraction of the main sequence lifetime.  
Since these stars are not expected to be spun down or up by any mass loss, the surface rotation velocity change is due to redistribution of angular momentum within the star.  The implication is that large natal shear has already been dissipated in these stars, and subsequent surface rotation evolution is due to the gradual decline of residual shear. A shear dynamo dissipates differential rotation, and the timescale over which this occurs is empirically limited by the shear energy available to power the observed X-rays.  \citet{Tout.Pringle:95} assumed a conversion efficiency of $10^{-3}$ for a timescale of about $10^6$~yr.  A longer timescale could be achieved assuming a greater conversion efficiency, but in addition to potential physical difficulties with a high efficiency of energy conversion, longer timescales of decay would not match either the slow evolution of surface rotation of most main-sequence A stars or  the observed plunge into X-ray darkness of early A stars following the pre-main sequence phase.

Difficulties facing shear-driven dynamo mechanisms of X-ray emission include recent theoretical work by \citet{Arlt.Rudiger:11}, who found only very 
weak dynamo action from numerical simulations of a 3D spherical shell with differential rotation, and potential difficulties in getting any dynamo-generated field to the stellar surface \citep[e.g.][]{Braithwaite.Cantiello:13}.    \citet{Hamaguchi.etal:05} suggest that X-ray activity is associated with accretion and jet activity, since the X-ray decline in Herbig~Ae stars seems to accompany the dissipation of  the accreting gas disk.  

The X-ray luminosity we predict from a subsurface He convection zone in late B and early A-type stars lies below our current upper limit for Vega by a factor of 2--3,  but carries a large uncertainty.  The lack of an X-ray detection of Vega then does provide some constraints to such a dynamo.  Pushing the X-ray detection limit significantly lower for Vega-like stars would be of considerable interest for investigating this further.

Detectable main sequence X-ray emission does appear to switch on at masses only slightly below that of Vega and HR4796A.  
Tantalizing evidence that young mid-A type stars can have residual X-ray 
activity came from the FUSE detection of transition region lines in the
spectrum of the 12~Myr old A6~V star $\beta$~Pic \citep{Bouret.etal:02}.  A
tentative detection of O~{\sc VII} emission was subsequently made from XMM-Newton
observations indicating  a plasma temperature of $6\times
10^5$~K \citep{Hempel.etal:05}.     
It was speculated that 
$\beta$~Pic either has a cool corona or a boundary layer between
the photosphere and its remnant disk \citep{Hempel.etal:05}.  A deep {\it Chandra} observation of $\beta$~Pic by \citet{Gunther.etal:12} has succeeded in detecting the star with $L_X=1.3\times 10^{27}$~erg~s$^{-1}$ in the 0.06--2 keV band.  The emission is consistent with that expected from an optically-thin plasma, and \citet{Gunther.etal:12} concluded the origin is coronal X-rays.  This makes $\beta$~Pic the hottest coronally active star detected so far.  $\beta$~Pic is also illustrated in Figure~\ref{f:toutpringle} and its position suggests the X-rays could originate from a shear dynamo.    However, 
 fairly cool ($3\times 10^6$~K) coronae have also been detected on the  planet-bearing A5~V star HR~8799 (aged about 60~Myr; \citealt{Robrade.Schmitt:10}) and on Altair (A7 IV-V; aged 1.2~Gyr; \citealt{Robrade.Schmitt:09}).  
 All three of these stars are near the boundary of both observational and theoretical temperature limits of significant photospheric 
 convection on A-type stars, and above which significant convection is not expected or observed \citep[e.g.][]{Landstreet.etal:09,Kupka.etal:09}.  The $\beta$~Pic, HR~8799 and Altair coronae then most likely originate from a convection-driven rather than shear-driven dynamo.

%

\section{Summary}


We have investigated the plight of X-ray emission in intermediate mass stars immediately following the Herbig Ae phase. 
A {\it Chandra} HRC-I observation of HR~4796A, an 8 Myr old main sequence A0 star devoid of close stellar companions, failed to detect the star, giving an upper limit to the X-ray luminosity of $1.3\times10^{27}$ erg s$^{-1}$.  This limit is still weakly consistent with predictions for dynamos driven by rotational shear and for an optimistic scenario of magnetic flux brought efficiently to the surface by magnetic buoyancy.  However, 
examining possible sources of X-rays from such stars in more detail, we find that the tapping of the large kinetic energy present in the star initially in the form of differential rotation is unlikely to produce observable X rays, chiefly because of the difficulty in getting the dissipated energy up to the surface of the star.   More promising is a subsurface convection layer produced by the ionisation of helium, which could host a dynamo. This mechanism, which should be effective throughout the main-sequence, could produce X ray luminosities of order $10^{25}$ erg s$^{-1}$---only moderately below the current detection limit for Vega.  It looks likely therefore that X-ray production in Herbig Ae/Be stars is linked to the accretion process rather than the properties of the star itself.

\acknowledgments

JJD and VK thank the NASA AISRP for providing financial assistance for the
development of the PINTofALE package.  This work was partially funded by {\it Chandra} grant GO7-8006X.
JJD and VK were funded by NASA contract
NAS8--03060 to the {\em Chandra X-ray Center} during the course of this
research and thank the Director, H.~Tananbaum, for continuing support
and encouragement.   Finally, we thank the referee, Christopher Tout, for insightful comments and suggestions that enabled us to significantly improve the manuscript.



\begin{thebibliography}{74}
\expandafter\ifx\csname natexlab\endcsname\relax\def\natexlab#1{#1}\fi

\bibitem[{{Arlt} \& {R{\"u}diger}(2011)}]{Arlt.Rudiger:11}
{Arlt}, R., \& {R{\"u}diger}, G. 2011, in IAU Symposium, Vol. 271, IAU
  Symposium, ed. {N.~H.~Brummell, A.~S.~Brun, M.~S.~Miesch, \& Y.~Ponty},
  213--217

\bibitem[{{Ayres}(2008)}]{Ayres:08}
{Ayres}, T.~R. 2008, \aj, 136, 1810

\bibitem[{{Barrado y Navascues}(1998)}]{Barrado_y_Navascues:98}
{Barrado y Navascues}, D. 1998, \aap, 339, 831

\bibitem[{{Bouret} {et~al.}(2002){Bouret}, {Deleuil}, {Lanz}, {Roberge},
  {Lecavelier des Etangs}, \& {Vidal-Madjar}}]{Bouret.etal:02}
{Bouret}, J.-C., {Deleuil}, M., {Lanz}, T., {Roberge}, A., {Lecavelier des
  Etangs}, A., \& {Vidal-Madjar}, A. 2002, \aap, 390, 1049

\bibitem[{{Braithwaite}(2006)}]{Braithwaite:06}
{Braithwaite}, J. 2006, \aap, 449, 451

\bibitem[{{Braithwaite}(2012)}]{Braithwaite:12}
---. 2012, \mnras, 422, 619

\bibitem[{{Braithwaite} \& {Cantiello}(2013)}]{Braithwaite.Cantiello:13}
{Braithwaite}, J., \& {Cantiello}, M. 2013, \mnras, 428, 2789

\bibitem[{{Brickhouse} {et~al.}(2010){Brickhouse}, {Cranmer}, {Dupree}, {Luna},
  \& {Wolk}}]{Brickhouse.etal:10}
{Brickhouse}, N.~S., {Cranmer}, S.~R., {Dupree}, A.~K., {Luna}, G.~J.~M., \&
  {Wolk}, S. 2010, \apj, 710, 1835

\bibitem[{{Cantiello} \& {Braithwaite}(2011)}]{Cantiello.Braithwaite:11}
{Cantiello}, M., \& {Braithwaite}, J. 2011, \aap, 534, A140

\bibitem[{{Damiani} {et~al.}(1994){Damiani}, {Micela}, {Sciortino}, \&
  {Harnden}}]{Damiani.etal:94}
{Damiani}, F., {Micela}, G., {Sciortino}, S., \& {Harnden}, Jr., F.~R. 1994,
  \apj, 436, 807

\bibitem[{{De Rosa} {et~al.}(2011){De Rosa}, {Bulger}, {Patience}, {Leland},
  {Macintosh}, {Schneider}, {Song}, {Marois}, {Graham}, {Bessell}, \&
  {Doyon}}]{De_Rosa.etal:11}
{De Rosa}, R.~J., {et~al.} 2011, \mnras, 415, 854

\bibitem[{{Drake}(2005)}]{Drake:05}
{Drake}, J.~J. 2005, in ESA Special Publication, Vol. 560, 13th Cambridge
  Workshop on Cool Stars, Stellar Systems and the Sun, ed. F.~{Favata},
  G.~A.~J. {Hussain}, \& B.~{Battrick}, 519--+

\bibitem[{{Drake}(1998)}]{Drake:98}
{Drake}, S.~A. 1998, Contributions of the Astronomical Observatory Skalnate
  Pleso, 27, 382

\bibitem[{{Grevesse} \& {Sauval}(1998)}]{Grevesse.Sauval:98}
{Grevesse}, N., \& {Sauval}, A.~J. 1998, Space Science Reviews, 85, 161

\bibitem[{{G{\"u}nther} {et~al.}(2006){G{\"u}nther}, {Liefke}, {Schmitt},
  {Robrade}, \& {Ness}}]{Gunther.etal:06}
{G{\"u}nther}, H.~M., {Liefke}, C., {Schmitt}, J.~H.~M.~M., {Robrade}, J., \&
  {Ness}, J.-U. 2006, \aap, 459, L29

\bibitem[{{G{\"u}nther} \& {Schmitt}(2009)}]{Gunther.Schmitt:09}
{G{\"u}nther}, H.~M., \& {Schmitt}, J.~H.~M.~M. 2009, \aap, 494, 1041

\bibitem[{{G{\"u}nther} {et~al.}(2012){G{\"u}nther}, {Wolk}, {Drake}, {Lisse},
  {Robrade}, \& {Schmitt}}]{Gunther.etal:12}
{G{\"u}nther}, H.~M., {Wolk}, S.~J., {Drake}, J.~J., {Lisse}, C.~M., {Robrade},
  J., \& {Schmitt}, J.~H.~M.~M. 2012, \apj, 750, 78

\bibitem[{{Hamaguchi} {et~al.}(2005){Hamaguchi}, {Yamauchi}, \&
  {Koyama}}]{Hamaguchi.etal:05}
{Hamaguchi}, K., {Yamauchi}, S., \& {Koyama}, K. 2005, \apj, 618, 360

\bibitem[{{Hamidouche} {et~al.}(2008){Hamidouche}, {Wang}, \&
  {Looney}}]{Hamidouche.etal:08}
{Hamidouche}, M., {Wang}, S., \& {Looney}, L.~W. 2008, \aj, 135, 1474

\bibitem[{{Hempel} {et~al.}(2005){Hempel}, {Robrade}, {Ness}, \&
  {Schmitt}}]{Hempel.etal:05}
{Hempel}, M., {Robrade}, J., {Ness}, J.-U., \& {Schmitt}, J.~H.~M.~M. 2005,
  \aap, 440, 727

\bibitem[{{Hill} {et~al.}(2010){Hill}, {Gulliver}, \& {Adelman}}]{Hill.etal:10}
{Hill}, G., {Gulliver}, A.~F., \& {Adelman}, S.~J. 2010, \apj, 712, 250

\bibitem[{{Hoffleit} \& {Jaschek}(1991)}]{Hoffleit.Jaschek:91}
{Hoffleit}, D., \& {Jaschek}, C.~. 1991, {The Bright star catalogue}, ed.
  {Hoffleit, D.~\& Jaschek, C.~|.}

\bibitem[{{Hubrig} {et~al.}(2014){Hubrig}, {Ilyin}, {Sch{\"o}ller}, {Cowley},
  {Castelli}, {Stelzer}, {Gonzalez}, \& {Wolff}}]{Hubrig.etal:14}
{Hubrig}, S., {Ilyin}, I., {Sch{\"o}ller}, M., {Cowley}, C.~R., {Castelli}, F.,
  {Stelzer}, B., {Gonzalez}, J.-F., \& {Wolff}, B. 2014, in European Physical
  Journal Web of Conferences, Vol.~64, European Physical Journal Web of
  Conferences, 8006

\bibitem[{{Jayawardhana} {et~al.}(1998){Jayawardhana}, {Fisher}, {Hartmann},
  {Telesco}, {Pina}, \& {Fazio}}]{Jayawardhana.etal:98}
{Jayawardhana}, R., {Fisher}, R.~S., {Hartmann}, L., {Telesco}, C., {Pina}, R.,
  \& {Fazio}, G. 1998, \apjl, 503, L79+

\bibitem[{{Jura}(1991)}]{Jura:91}
{Jura}, M. 1991, \apjl, 383, L79+

\bibitem[{{Jura} {et~al.}(1998){Jura}, {Malkan}, {White}, {Telesco}, {Pina}, \&
  {Fisher}}]{Jura.etal:98}
{Jura}, M., {Malkan}, M., {White}, R., {Telesco}, C., {Pina}, R., \& {Fisher},
  R.~S. 1998, \apj, 505, 897

\bibitem[{{Kashyap} \& {Drake}(2000)}]{Kashyap.Drake:00}
{Kashyap}, V., \& {Drake}, J.~J. 2000, Bulletin of the Astronomical Society of
  India, 28, 475

\bibitem[{{Kashyap} {et~al.}(2010){Kashyap}, {van Dyk}, {Connors}, {Freeman},
  {Siemiginowska}, {Xu}, \& {Zezas}}]{Kashyap.etal:10}
{Kashyap}, V.~L., {van Dyk}, D.~A., {Connors}, A., {Freeman}, P.~E.,
  {Siemiginowska}, A., {Xu}, J., \& {Zezas}, A. 2010, \apj, 719, 900

\bibitem[{{Kastner} {et~al.}(2002){Kastner}, {Huenemoerder}, {Schulz},
  {Canizares}, \& {Weintraub}}]{Kastner.etal:02}
{Kastner}, J.~H., {Huenemoerder}, D.~P., {Schulz}, N.~S., {Canizares}, C.~R.,
  \& {Weintraub}, D.~A. 2002, \apj, 567, 434

\bibitem[{{Koerner} {et~al.}(1998){Koerner}, {Ressler}, {Werner}, \&
  {Backman}}]{Koerner.etal:98}
{Koerner}, D.~W., {Ressler}, M.~E., {Werner}, M.~W., \& {Backman}, D.~E. 1998,
  \apjl, 503, L83+

\bibitem[{{Kupka} {et~al.}(2009){Kupka}, {Ballot}, \&
  {Muthsam}}]{Kupka.etal:09}
{Kupka}, F., {Ballot}, J., \& {Muthsam}, H.~J. 2009, Communications in
  Asteroseismology, 160, 30

\bibitem[{{Landstreet} {et~al.}(2009){Landstreet}, {Kupka}, {Ford}, {Officer},
  {Sigut}, {Silaj}, {Strasser}, \& {Townshend}}]{Landstreet.etal:09}
{Landstreet}, J.~D., {Kupka}, F., {Ford}, H.~A., {Officer}, T., {Sigut},
  T.~A.~A., {Silaj}, J., {Strasser}, S., \& {Townshend}, A. 2009, \aap, 503,
  973

\bibitem[{{MacGregor} \& {Cassinelli}(2003)}]{MacGregor.Cassinelli:03}
{MacGregor}, K.~B., \& {Cassinelli}, J.~P. 2003, \apj, 586, 480

\bibitem[{{Malkov}(2007)}]{Malkov:07}
{Malkov}, O.~Y. 2007, \mnras, 382, 1073

\bibitem[{{Mullan} \& {MacDonald}(2005)}]{Mullan.MacDonald:05}
{Mullan}, D.~J., \& {MacDonald}, J. 2005, \mnras, 356, 1139

\bibitem[{{Parker}(1966)}]{Parker:66}
{Parker}, E.~N. 1966, \apj, 143, 32

\bibitem[{{Pease} {et~al.}(2006){Pease}, {Drake}, \& {Kashyap}}]{Pease.etal:06}
{Pease}, D.~O., {Drake}, J.~J., \& {Kashyap}, V.~L. 2006, \apj, 636, 426

\bibitem[{{Pizzolato} {et~al.}(2003){Pizzolato}, {Maggio}, {Micela},
  {Sciortino}, \& {Ventura}}]{Pizzolato.etal:03}
{Pizzolato}, N., {Maggio}, A., {Micela}, G., {Sciortino}, S., \& {Ventura}, P.
  2003, \aap, 397, 147

\bibitem[{{Rhee} {et~al.}(2007){Rhee}, {Song}, \& {Zuckerman}}]{Rhee.etal:07}
{Rhee}, J.~H., {Song}, I., \& {Zuckerman}, B. 2007, \apj, 671, 616

\bibitem[{{Robrade} \& {Schmitt}(2009)}]{Robrade.Schmitt:09}
{Robrade}, J., \& {Schmitt}, J.~H.~M.~M. 2009, \aap, 497, 511

\bibitem[{{Robrade} \& {Schmitt}(2010)}]{Robrade.Schmitt:10}
---. 2010, \aap, 516, A38

\bibitem[{{Robrade} \& {Schmitt}(2011)}]{Robrade.Schmitt:11}
---. 2011, \aap, 531, A58

\bibitem[{{Royer} {et~al.}(2002){Royer}, {Gerbaldi}, {Faraggiana}, \&
  {G{\'o}mez}}]{Royer.etal:02}
{Royer}, F., {Gerbaldi}, M., {Faraggiana}, R., \& {G{\'o}mez}, A.~E. 2002,
  \aap, 381, 105

\bibitem[{{Schneider} {et~al.}(2009){Schneider}, {Weinberger}, {Becklin},
  {Debes}, \& {Smith}}]{Schneider.etal:09}
{Schneider}, G., {Weinberger}, A.~J., {Becklin}, E.~E., {Debes}, J.~H., \&
  {Smith}, B.~A. 2009, \aj, 137, 53

\bibitem[{{Schneider} {et~al.}(1999){Schneider}, {Smith}, {Becklin}, {Koerner},
  {Meier}, {Hines}, {Lowrance}, {Terrile}, {Thompson}, \&
  {Rieke}}]{Schneider.etal:99}
{Schneider}, G., {et~al.} 1999, \apjl, 513, L127

\bibitem[{{Schr{\"o}der} \& {Schmitt}(2007)}]{Schroder.Schmitt:07}
{Schr{\"o}der}, C., \& {Schmitt}, J.~H.~M.~M. 2007, \aap, 475, 677

\bibitem[{{Simon} {et~al.}(1995){Simon}, {Drake}, \& {Kim}}]{Simon.etal:95}
{Simon}, T., {Drake}, S.~A., \& {Kim}, P.~D. 1995, \pasp, 107, 1034

\bibitem[{{Skinner} {et~al.}(2004){Skinner}, {G{\"u}del}, {Audard}, \&
  {Smith}}]{Skinner.etal:04}
{Skinner}, S.~L., {G{\"u}del}, M., {Audard}, M., \& {Smith}, K. 2004, \apj,
  614, 221

\bibitem[{{Smith} {et~al.}(2001){Smith}, {Brickhouse}, {Liedahl}, \&
  {Raymond}}]{Smith.etal:01}
{Smith}, R.~K., {Brickhouse}, N.~S., {Liedahl}, D.~A., \& {Raymond}, J.~C.
  2001, \apjl, 556, L91

\bibitem[{{Soderblom} {et~al.}(2013){Soderblom}, {Hillenbrand}, {Jeffries},
  {Mamajek}, \& {Naylor}}]{Soderblom.etal:13}
{Soderblom}, D.~R., {Hillenbrand}, L.~A., {Jeffries}, R.~D., {Mamajek}, E.~E.,
  \& {Naylor}, T. 2013, ArXiv e-prints

\bibitem[{{Spruit}(1999)}]{Spruit:99}
{Spruit}, H.~C. 1999, \aap, 349, 189

\bibitem[{{Spruit}(2002)}]{Spruit:02}
---. 2002, \aap, 381, 923

\bibitem[{{Stauffer} {et~al.}(1995){Stauffer}, {Hartmann}, \& {Barrado y
  Navascues}}]{Stauffer.etal:95}
{Stauffer}, J.~R., {Hartmann}, L.~W., \& {Barrado y Navascues}, D. 1995, \apj,
  454, 910

\bibitem[{{Stelzer} {et~al.}(2011){Stelzer}, {Hummel}, {Sch{\"o}ller},
  {Hubrig}, \& {Cowley}}]{Stelzer.etal:11}
{Stelzer}, B., {Hummel}, C.~A., {Sch{\"o}ller}, M., {Hubrig}, S., \& {Cowley},
  C. 2011, \aap, 529, A29

\bibitem[{{Stelzer} {et~al.}(2006){Stelzer}, {Micela}, {Hamaguchi}, \&
  {Schmitt}}]{Stelzer.etal:06}
{Stelzer}, B., {Micela}, G., {Hamaguchi}, K., \& {Schmitt}, J.~H.~M.~M. 2006,
  \aap, 457, 223

\bibitem[{{Stelzer} {et~al.}(2009){Stelzer}, {Robrade}, {Schmitt}, \&
  {Bouvier}}]{Stelzer.etal:09}
{Stelzer}, B., {Robrade}, J., {Schmitt}, J.~H.~M.~M., \& {Bouvier}, J. 2009,
  \aap, 493, 1109

\bibitem[{{Stelzer} \& {Schmitt}(2004)}]{Stelzer.Schmitt:04}
{Stelzer}, B., \& {Schmitt}, J.~H.~M.~M. 2004, \aap, 418, 687

\bibitem[{{Swartz} {et~al.}(2005){Swartz}, {Drake}, {Elsner}, {Ghosh}, {Grady},
  {Wassell}, {Woodgate}, \& {Kimble}}]{Swartz.etal:05}
{Swartz}, D.~A., {Drake}, J.~J., {Elsner}, R.~F., {Ghosh}, K.~K., {Grady},
  C.~A., {Wassell}, E., {Woodgate}, B.~E., \& {Kimble}, R.~A. 2005, \apj, 628,
  811

\bibitem[{{Tayler}(1973)}]{Tayler:73}
{Tayler}, R.~J. 1973, \mnras, 161, 365

\bibitem[{{Thalmann} {et~al.}(2011){Thalmann}, {Janson}, {Buenzli}, {Brandt},
  {Wisniewski}, {Moro-Mart{\'{\i}}n}, {Usuda}, {Schneider}, {Carson},
  {McElwain}, {Grady}, {Goto}, {Abe}, {Brandner}, {Dominik}, {Egner}, {Feldt},
  {Fukue}, {Golota}, {Guyon}, {Hashimoto}, {Hayano}, {Hayashi}, {Hayashi},
  {Henning}, {Hodapp}, {Ishii}, {Iye}, {Kandori}, {Knapp}, {Kudo}, {Kusakabe},
  {Kuzuhara}, {Matsuo}, {Miyama}, {Morino}, {Nishimura}, {Pyo}, {Serabyn},
  {Suto}, {Suzuki}, {Takahashi}, {Takami}, {Takato}, {Terada}, {Tomono},
  {Turner}, {Watanabe}, {Yamada}, {Takami}, \& {Tamura}}]{Thalmann.etal:11}
{Thalmann}, C., {et~al.} 2011, \apjl, 743, L6

\bibitem[{{Tout} {et~al.}(1999){Tout}, {Livio}, \& {Bonnell}}]{Tout.etal:99}
{Tout}, C.~A., {Livio}, M., \& {Bonnell}, I.~A. 1999, \mnras, 310, 360

\bibitem[{{Tout} \& {Pringle}(1992)}]{Tout.Pringle:92}
{Tout}, C.~A., \& {Pringle}, J.~E. 1992, \mnras, 256, 269

\bibitem[{{Tout} \& {Pringle}(1995)}]{Tout.Pringle:95}
---. 1995, \mnras, 272, 528

\bibitem[{{Vaiana} {et~al.}(1981){Vaiana}, {Cassinelli}, {Fabbiano},
  {Giacconi}, {Golub}, {Gorenstein}, {Haisch}, {Harnden}, {Johnson}, {Linsky},
  {Maxson}, {Mewe}, {Rosner}, {Seward}, {Topka}, \& {Zwaan}}]{Vaiana.etal:81}
{Vaiana}, G.~S., {et~al.} 1981, \apj, 245, 163

\bibitem[{{van Dyk} {et~al.}(2001){van Dyk}, {Connors}, {Kashyap}, \&
  {Siemiginowska}}]{van_Dyk.etal:01}
{van Dyk}, D.~A., {Connors}, A., {Kashyap}, V.~L., \& {Siemiginowska}, A. 2001,
  \apj, 548, 224

\bibitem[{{van Leeuwen}(2007)}]{van_Leeuwen:07}
{van Leeuwen}, F. 2007, \aap, 474, 653

\bibitem[{{Vigneron} {et~al.}(1990){Vigneron}, {Mangeney}, {Catala}, \&
  {Schatzman}}]{Vigneron.etal:90}
{Vigneron}, C., {Mangeney}, A., {Catala}, C., \& {Schatzman}, E. 1990,
  \solphys, 128, 287

\bibitem[{{Webb} {et~al.}(1999){Webb}, {Zuckerman}, {Platais}, {Patience},
  {White}, {Schwartz}, \& {McCarthy}}]{Webb.etal:99}
{Webb}, R.~A., {Zuckerman}, B., {Platais}, I., {Patience}, J., {White}, R.~J.,
  {Schwartz}, M.~J., \& {McCarthy}, C. 1999, \apjl, 512, L63

\bibitem[{{Weinberger} {et~al.}(2013){Weinberger}, {Anglada-Escud{\'e}}, \&
  {Boss}}]{Weinberger.etal:13}
{Weinberger}, A.~J., {Anglada-Escud{\'e}}, G., \& {Boss}, A.~P. 2013, \apj,
  767, 96

\bibitem[{{Wright} {et~al.}(2011){Wright}, {Drake}, {Mamajek}, \&
  {Henry}}]{Wright.etal:11}
{Wright}, N.~J., {Drake}, J.~J., {Mamajek}, E.~E., \& {Henry}, G.~W. 2011,
  \apj, 743, 48

\bibitem[{{Wyatt} {et~al.}(1999){Wyatt}, {Dermott}, {Telesco}, {Fisher},
  {Grogan}, {Holmes}, \& {Pi{\~n}a}}]{Wyatt.etal:99}
{Wyatt}, M.~C., {Dermott}, S.~F., {Telesco}, C.~M., {Fisher}, R.~S., {Grogan},
  K., {Holmes}, E.~K., \& {Pi{\~n}a}, R.~K. 1999, \apj, 527, 918

\bibitem[{{Yoon} {et~al.}(2010){Yoon}, {Peterson}, {Kurucz}, \&
  {Zagarello}}]{Yoon.etal:10}
{Yoon}, J., {Peterson}, D.~M., {Kurucz}, R.~L., \& {Zagarello}, R.~J. 2010,
  \apj, 708, 71

\bibitem[{{Zinnecker} \& {Preibisch}(1994)}]{Zinnecker.Preibisch:94}
{Zinnecker}, H., \& {Preibisch}, T. 1994, \aap, 292, 152

\bibitem[{{Zorec} \& {Royer}(2012)}]{Zorec.Royer:12}
{Zorec}, J., \& {Royer}, F. 2012, \aap, 537, A120

\end{thebibliography}

\end{document}